\documentclass[12pt]{article}
\usepackage{eurosym}
\usepackage[left=0.5in, right=0.5in, top=0.75in, bottom=0.75in]{geometry}
\usepackage{hyperref}
\usepackage{mathtools}
\usepackage{graphicx}
\usepackage{float}
\usepackage{blindtext}
\usepackage{mathrsfs}
\usepackage{pstricks}
\usepackage{amsmath}
\usepackage{multirow}
\usepackage{booktabs,tabularx}
\newcolumntype{C}{>{\centering\arraybackslash}X}
\usepackage[figurename=Fig.]{caption}

\newcommand{\bra}[1]{\bigl\langle #1 \bigr|}
\newcommand{\ket}[1]{\bigl| #1 \bigr\rangle}
\begin{document}

	\renewcommand{\baselinestretch}{1.3} \topmargin=-1.8cm \textheight=23 cm
	\textwidth=23cm
	
	\begin{center}
		          \textit{\Large   Wigner  distribution function of atomic system interacts locally with a deformed cavity \\ }
			\bigskip	
	\textit{M. Y. Abd-Rabbou} $^{a}$ \textit{\footnote{e-mail:m.elmalky@azhar.edu.eg}},\textit{N. Metwally}$^{b,c}$\textit{\footnote{nmetwally@gmail.com}}, \textit{M. M. A. Ahmed}$^{a}$, and \textit{A.-S. F. Obada}$^{a}$ ,

		 $^{a}${\footnotesize Mathematics Department, Faculty of Science, Al-Azhar
		 	University, Nasr City 11884, Cairo, Egypt.}
		
		 $^{b}${\footnotesize Math. Dept., College of Science, University of Bahrain, Bahrain.}
		
		 $^{c}${\footnotesize Department of Mathematics, Aswan University
		 	Aswan, Sahari 81528, Egypt.}
	
	\end{center}
\begin{abstract}

 Wigner distribution function  of  atomic system interacts locally with a deformed cavity is discussed.
It is shown that, the deformed cavity has a destructive effect on the Wigner distribution
function, where it decreases as one increases the deformation strength.  The upper and lower bounds of the Wigner distribution function depends on the initial state settings of atomic system (entangled/product), the initial values of the dipole-dipole interaction's and detuning  parameters, and the external distribution weight and the phase angles.  The possibility of suppressing the decay induced by the deformed cavity
may be increased by increasing the dipole's strength or the detuning parameter.  We show that the distribution angles may be considered as a control external parameters, that maximize/ minimize the Wigner distribution function. This means that by controlling on the distribution angles, one can increase the possibility of suppressing the decoherence induced by the deformed cavity.

\end{abstract}
\section{Introduction.}
  There are several studies have devoted to investigate  the treatment of   information between two users, who share different systems. Different studies have introduced to discuss many physical properties of atomic systems interact with a quantized cavity mode field \cite{abdalla2015quantum}.
  Entanglement as a non-local property that generated between  atoms inside different types of cavity modes is investigated for different systems\cite{hessian2011entanglement}. The purity and the fidelity via deformed cavity field interacted with a pair of entangled qubits is discussed \cite{metwally2011dynamics}. The  information of the  2-qubit system is investigated within the magnetic field, non-Markovian environments, and acceleration \cite{ffff,franco2013dynamics,abd2019wigner}.
On the other hand, the $q$-deformed of the Heisenberg algebra has many physical applications  \cite{lavagno2008deformed,cerchiai1999geometry,h2004dynamical,haghshenasfard2012collective,othman2018interaction}.  Lavagno\cite{lavagno2008deformed} has obtained a generalized differential form of linear Schr\"{o}dinger equation which involves the  $q$-deformed Hamiltonian that is non-Hermitian. The geometry of the $q$-deformed phase space has investigated by Cerchiai, et. al \cite{cerchiai1999geometry}. Naderi et. al \cite{h2004dynamical} have studied the temporal evolution of the atomic population inversion and quantum fluctuations of  the two-photon $q$-deformed Jaynes-Cummings model. However, the framework of a one and two-photon $q$-deformed Dicke model by using a Bose-Einstein condensate is investigated in \cite{haghshenasfard2012collective}. Meanwhile, the interaction between the deformed electromagnetic field and $N$-type four level atoms in the presence of a nonlinear Kerr medium is discussed\cite{othman2018interaction}.

As far as we know, the Wigner function play significant role in quantum distribution theory, where it indicates the quantumness analog by its negative values.  Also, the Wigner function depicts the classicality by its positive values\cite{gggg,abd2019external}. For the  atomic systems  the Wigner function has not been studied widely. Among of these studies, for a two-qubit interacting with a quantized field system the Wigner distribution function is discussed analytically in the presence of pure phase noisy channel\cite{obada2012wigner}. However,  the Wigner function has been  investigated the phase space for the quantum states based on finite fields with continuous or discrete degrees of freedom  \cite{PhysRevA.70.062101}. The time evolution of Wigner function in  SU(2) algebra  is analyzed for superconducting flux qubits \cite{reboiro2015use}. Also, It has been reconstructed in SU(2) algebra for the accelerated three qubit system in the presence of some noisy channels \cite{abd2019wigner1}.

In this manuscript, we are motivated to study the behavior of the Wigner distribution function of the atomic system consisted  of two atoms. It is assumed that the atomic system either prepared in a maximum entangled or product state. This atomic system interacts locally with a cavity mode described ny imperfect (deformed) operators. We investigate the effect of the field and the atomic system's parameters on the behavior of the Wigner distribution function. Moreover, the distribution angles are considered as an external parameter that can be used to maximize/ minimize the Wigner function distribution.

  Therefore, the paper is organized this  as : Sec. \ref{s6.2} is devoted to present the suggested physical model and demonstrate the exact solution of the evolution system.  We review the mathematical form of the Wigner function in Sec.\ref{s6.3}.  Finally, our results are summarized in Sec.\ref{s6.4}.

\section{The suggest physical model.}\label{s6.2}
 Let us assume that the  two users, Alice and Bob   share an atomic system consists of 2-two level atoms (two-qubits), is prepared initially in a partial entangled state defined as,

 \begin{eqnarray}
	|\psi_{ab}(0)\rangle&=&a_1|1,1\rangle+a_2|1,0\rangle+a_3|0,1\rangle+a_4|0,0\rangle.
\end{eqnarray}
where $|a_1^2+a_2^2+a_3^2+a_4^2=1$. The initial entanglement of this state depends on the choices of these coefficients. The two atoms may be prepared  in an excited state if we set  $a_1=1, a_2=0=a_3=a_4$. The  maximum entangled state of Bell types can be obtained, if we set $a_1=a_2=\frac{1}{\sqrt{2}}$ and $a_3=a_2=0$. Otherwise, one can obtain a partial entangled/ separable state.  It is assumed that, this atomic system interacts locally with a cavity deformed mode is initially prepared in the coherent state,
 \begin{eqnarray}
	|\psi_f(0)\rangle&=&\sum_{n=0}{}q_n|n\rangle,\ q_n=\frac{\alpha^n}{\sqrt{n}}e^{\frac{-|\alpha|^2}{2}},
\end{eqnarray}
Therefore,  initial state of the atomic-field system is given by,
\begin{equation}\label{InS}
\ket{\psi_s(0)}=\ket{\psi_{ab}(0)}\otimes\ket{\psi_f(0)}.
\end{equation}
In the  rotating-wave approximations, the   Hamiltonian which describes the atomic-field system may be written as:
\begin{equation}\label{6.1}
	\hat{H}=\omega_f \hat{\mathcal{R}}^{\dagger}\hat{\mathcal{R}}+  \sum_{i=1}^{2} \big(\Omega_i\hat{\sigma}_z^{(i)}+\lambda (\hat{\mathcal{R}}\hat{\sigma}_{+}^{(i)}+\hat{\mathcal{R}}^{\dagger}\hat{\sigma}_{-}^{(i)})\big)+i \kappa_d (\hat{\sigma}_{+}^{(1)} \hat{\sigma}_{-}^{(2)}- \hat{\sigma}_{-}^{(1)} \hat{\sigma}_{+}^{(2)}),
\end{equation}
where $ \omega_f $ and $ \Omega_i $ are  the frequencies  of deformed field and the atomic system, respectively. The parameter  $ \lambda $  represents the coupling constant between  the deformed field and  the two  atoms,  while the  parameter $ g $ is the coupling constant between the two atoms.
 The operators $\hat{\sigma}_{z}^{(i)} $, $ \hat{\sigma}_{+}^{(i)} $, and $ \hat{\sigma}_{-}^{(i)} $ are the Pauli spin operators. Finally, the operators  $ \hat{\mathcal{R}}^{\dagger} $ and $ \hat{\mathcal{R}} $ are the creation and annihilation  operators of the  generalized deformation field, which they  are defined by well known  bosonic  operators $ \hat{a}^{\dagger}$ and $ \hat{a}  $ as:
\begin{equation}
	\hat{\mathcal{R}}^{\dagger}=\hat{a}^{\dagger} f(\hat{n}+1), \qquad \hat{\mathcal{R}}=\hat{a} f(\hat{n}), \quad \text{with} \quad \hat{n}= \hat{a}^{\dagger} \hat{a}.
\end{equation}
These operators satisfy the commutation relation,
\begin{equation}
	[\hat{\mathcal{R}},\hat{\mathcal{R}}^{\dagger}]=(\hat{n}+1)f^2(\hat{n}+1)+\hat{n} f^2(\hat{n}),\quad \text{with} \qquad [\hat{a},\hat{a}^{\dagger}]=I.
\end{equation}
The function $f(n)$ is  an arbitrary function represents the deformation function. In this contribution we consider the q-deformation \cite{arik1976hilbert}, which is  represented in terms of the $ q $ parameter as,
\begin{equation}
f(n)=\sqrt{\frac{1-q^n}{n(1-q)}}.
\end{equation}

In the Heisenberg  picture, the equations of motion  of the suggested system are given by means of the operators, $ \hat{\mathcal{R}} $, $ \hat{\mathcal{R}}^{\dagger}  $ and $ \hat{\sigma}_{z}^{(i)} $ as,
\begin{eqnarray}\label{Heis}
	\frac{d \hat{\mathcal{R}}}{d t}&=& - i \omega_f \hat{\mathcal{R}} - i \lambda \sum_{i=1}^{2} \hat{\sigma}_{-}^{(i)}, \qquad \frac{d \hat{\mathcal{R}}^{\dagger}}{d t}= i \omega_f \hat{\mathcal{R}}^{\dagger}+ i \lambda \sum_{i=1}^{2} \hat{\sigma}_{+}^{(i)}, \nonumber\\ \frac{d \hat{\sigma}_{z}^{(1)}}{d t}&=& 2 i	\lambda (\hat{\mathcal{R}} \hat{\sigma}_{+}^{(1)}+ \hat{\mathcal{R}}^{\dagger} \hat{\sigma}_{-}^{(1)})-2 \kappa_d \sum_{i\neq j=1}^{2} \hat{\sigma}_{+}^{(i)} \hat{\sigma}_{-}^{(j)}, \nonumber\\
	 \frac{d \hat{\sigma}_{z}^{(2)}}{d t}&=& 2 i \lambda (\hat{\mathcal{R}} \hat{\sigma}_{+}^{(2)}+ \hat{\mathcal{R}}^{\dagger} \hat{\sigma}_{-}^{(2)})+2 \kappa_d \sum_{i\neq j=1}^{2} \hat{\sigma}_{+}^{(i)} \hat{\sigma}_{-}^{(j)}.
\end{eqnarray}
The system of equations (\ref{Heis} may be used to describe the  Hamiltonian model in eq.(\ref{6.1}) in terms of constant of motion as:
\begin{equation}
	\frac{\hat{H}}{\hbar}=\omega_f \hat{\mathcal{H}}_0+ \hat{\mathcal{H}}_I, \quad \text{where} \quad \hat{\mathcal{H}}_0= \hat{\mathcal{R}}^{\dagger}\hat{\mathcal{R}}+\frac{1}{2} \sum_{i=1}^{2} \hat{\sigma}_{z}^{(i)},
\end{equation}
while  the interaction Hamiltonian operator $\hat{\mathcal{H}}_I$ is;
\begin{equation}\label{6.6}
	\hat{\mathcal{H}}_I= \sum_{i=1}^{2} \big(\frac{\Delta_i}{2}  \hat{\sigma}_z^{(i)}+\lambda (\hat{\mathcal{R}}\hat{\sigma}_{+}^{(i)}+\hat{\mathcal{R}}^{\dagger}\hat{\sigma}_{-}^{(i)})\big)+i\kappa_d(\hat{\sigma}_{+}^{(1)} \hat{\sigma}_{-}^{(2)}- \hat{\sigma}_{-}^{(1)} \hat{\sigma}_{+}^{(2)}),\quad
\end{equation}
where the detuning parameter, $\Delta_i=2\Omega_i-\omega_f, \quad i=1,2$.
At any $ t>0 $ the time evaluation of the initial atomic-field system  (\ref{InS}) may be written as,
\begin{equation}\label{6.7}
	|\psi_s(t)\rangle=\sum_{n=0}^{\infty}(C^n_1(t)|1,1,n\rangle+C_2^n(t)|1,0,n+1\rangle+C^n_3(t)|0,1,n+1\rangle+C^n_4(t)|0,0,n+2\rangle),
\end{equation}
where $C_i, i=1,..4$, with $ \sum_{j=1}^{4}|C_j|^2=1$ are the solutions of the following system of differential equations,
\begin{equation}\label{6.8}
\begin{bmatrix} \partial_t C^n_1(t)\\ \partial_t C^n_2(t)\\\partial_t C^n_3(t)\\\partial_t C^n_4(t) \end{bmatrix}=-i \begin{bmatrix}
0 & \nu_1(n) & \nu_1(n)& 0\\\nu_1(n)  & \delta & i \kappa_d & \nu_2(n) \\\nu_1(n)  & -\delta & -i \kappa_d & \nu_2(n)\\ 0 & \nu_2(n) & \nu_2(n)& 0 \end{bmatrix} \begin{bmatrix}
C^n_1(t)\\ C^n_2(t)\\ C^n_3(t)\\ C^n_4(t)
\end{bmatrix}
\end{equation}
where $\nu_1(n)=\lambda f(n+1)\sqrt{n+1}  $, $\nu_2(n)=\lambda f(n+2)\sqrt{n+2} $, and $ \Delta_1-\Delta_2=2\delta $, where we simulating that the atomic system in an exact resonance case, i.e. $ \Delta_1+\Delta_2=0 $.  The solution of this system is obtained analytically as,
\begin{eqnarray}\label{Ampl}
C^n_1(t)&=&q_n a_1-\frac{K_1}{\mu^2}(1-\cos \mu t)-\frac{i \nu_1(n) q_{n+1}(a_2+a_3)}{\mu}\sin \mu t,\nonumber\\
C^n_2(t)&=&q_{n+1} a_2 \cos \mu t -\frac{\chi}{\mu^2}(1-\cos \mu t) -\frac{i G_1}{\mu}\sin \mu t,\nonumber\\
C^n_3(t)&=& q_{n+1} a_3 \cos \mu t +\frac{\chi}{\mu^2}(1-\cos \mu t) -\frac{i G_2}{\mu}\sin \mu t,\nonumber\\
C^n_4(t)&=&q_{n+2} a_4-\frac{K_2}{\mu^2}(1-\cos \mu t)-\frac{i \nu_2(n) q_{n+1}(a_2+a_3)}{\mu}\sin \mu t,
\end{eqnarray}
with
\begin{eqnarray}
	K_j&=&\nu_j(n) q_{n+1}(a_2-a_3)(\delta-i \kappa_d)+2 \nu_j(n) (q_{n} a_1 \nu_1(n)+q_{n+1} a_2\nu_2(n)), \quad j=1,2\nonumber\\
	\chi&=&(q_{n} a_1 \nu_1(n)+q_{n+2} a_4\nu_2(n))(\delta-i \kappa_d)-q_{n+1} (a_2-a_3)(\nu_1^2(n)+\nu_1^2(n)),\nonumber\\
	G_1&=&q_n a_1 \nu_1(n)+q_{n+1} a_2 \delta+i q_{n+1} a_3 \kappa_d+q_{n+2} a_4,\nonumber\\
	G_2&=&q_n a_1 \nu_1(n)-i q_{n+1} a_2 \kappa_d - q_{n+1} a_3 \delta+q_{n+2} a_4,
\end{eqnarray}
where $\mu=\sqrt{\delta^2+2(\nu_1^2(n)+\nu_1^2(n))+\kappa_d^2}$. Now, by  using Eqs.(\ref{6.7} and (\ref{Ampl}) , the final state of the total systems can be written as,
\begin{equation}\label{dins}
\begin{split}
\hat{\rho_s}=|\psi_s(t)\rangle \langle \psi_s(t)|.
\end{split}
\end{equation}

  The main task of this manuscript is investigating the effect of the deformed field on the behavior of the Wigner distribution function.  Therefore, it is important to obtain the states of Alice-Bob atomic subsystems and the Alice subsystem. to obtain the atomic  state, one traces out the field degree of freedom, i.e. $\hat{\rho}_{A,B}=Tr_{field}|\psi(t)\rangle \langle \psi(t)|$. So, the  reduced density operator of Alice, and Bob atomic subsystem is given by,
\begin{equation}\label{DAB}
\begin{split}
\hat{\rho}_{A,B} &=\begin{bmatrix}
|C^n_1(t)|^2  & C^{n+1}_1(t) C^{n^*}_2(t) & C^{n+1}_1(t) C^{n^*}_3(t)& C^{n+2}_1(t) C^{n^*}_4(t)\\C^{n+1^*}_1(t) C^n_2(t)  & |C^{n1}_2(t)|^2  & C^{n}_2(t) C^{n^*}_3(t)& C^{n+1}_2(t) C^{n^*}_4(t) \\C^{n+1^*}_1(t) C^{n}_3(t) & C^{n^*}_2(t) C^{n}_3(t) & |C^{n}_3(t)|^2 & C^{n+1}_3(t) C^{n^*}_4 (t)\\  C^{n+2^*}_1(t) C^{n}_4(t) &  C^{n+2^*}_1(t) C^{n}_4(t)& C^{n+1^*}_3(t) C^{n}_4 (t)& |C^n_4(t)|^2 \end{bmatrix}
\end{split}
\end{equation}

On the other hand, one can obtain the  reduced density operator of Alice atomic subsystem as,
\begin{eqnarray}\label{DA}
\hat{\rho}_A &=&Tr_{B} [\hat{\rho}_{A,B}]
\nonumber\\&=&\big(|C^n_1(t)|^2+|C^n_2(t)|^2\big)|1\rangle_A\langle 1|+\big(|C^n_3(t)|^2+|C^n_4(t)|^2\big)|0\rangle_A\langle 0|\nonumber\\ &+ &\bigg(C^{n+1}_1(t) C^{n^*}_3(t)+C^{n+1}_21(t) C^{n^*}_4(t)\bigg)|1\rangle_A\langle 0|+h.c.
\end{eqnarray}

\section{Wigner Distribution Function.}\label{s6.3}

The atomic Wigner probability distribution in SU(2) algebra is reconstructed in the angular momentum basis of a two-level atom $ |m,\frac{1}{2}\rangle  , m=-\frac{1}{2},\frac{1}{2} $ as follows \cite{PhysRevA.24.2889,abd2019wigner}:
\begin{equation}\label{3.1}
\mathcal{W}_{\hat{\rho}}(\theta,\phi)=Tr[\hat{\rho}_{a,b}\hat{A}_a(\theta,\phi)\hat{A}_b(\theta,\phi)],
\end{equation}
where the  operator $ \hat{A}_i (\theta,\phi)$, $ i=a,b $ corresponds to the two subsystem Alice and Bob is defined by:
\begin{equation}\label{3.2}
\hat{A}_i(\theta,\phi)=\sqrt{2\pi} \sum_{L_i=0}^{1} \sum_{M=-L}^{L} \hat{T}^{(i)^{\dagger}}_{L,M} Y^i _{L,M}(\theta,\phi),
\end{equation}
The  $Y^i_{L,M}(\theta,\phi) $ are the spherical harmonics functions, while $ T^{(i)^{\dagger}}_{L,M}= (-1)^{M}  T_{L,-M}$ are the orthogonal irreducible tensor operators which  are represented in Hilbert space as a linear combination by \cite{klimov2017generalized}:

\begin{equation}\label{3.3}
\hat{T}^{(i)^{\dagger}}_{L,M}=(-1)^{M}\sqrt{\frac{2L+1}{2}} \sum_{m,m'=-\frac{1}{2}}^{\frac{1}{2}} C^{\frac{1}{2}, m'}_{\frac{1}{2},m;L,-M}|\frac{1}{2},m'\rangle\langle \frac{1}{2},m |.
\end{equation}
The coefficient $ C^{\frac{1}{2},m'}_{\frac{1}{2},m;L,-M} $ is the Clebsch-Gordan coupling coefficient, where $ 0 \leq L_i \leq 1 $, and $ -L\leq M \leq L $.
Then, the Wigner probability distribution is given by,
\begin{equation}\label{3.4}
\begin{split}
\mathcal{W}_{\hat{\rho}}(\theta,\phi)=2\pi \ Tr&\bigg[\hat{\rho}_{AB}\prod_{i}^{n}\big( \hat{T}^{i^{\dagger}}_{0,0} Y^i_{0,0}(\theta,\phi)+ \sum_{n=-1}^{1} \hat{T}^{i^{\dagger}}_{1,n} Y^i_{1,n}(\theta,\phi)\big)\bigg],
\end{split}
\end{equation}
where,
\begin{equation*}
\begin{split}
&\hat{T}^{(i)^{\dagger}}_{0,0}=\frac{1}{\sqrt{2}}(|0\rangle_i \langle 0|+ |1\rangle_i \langle 1|), \quad \hat{T}^{(i)^{\dagger}}_{1,0}=\frac{-1}{\sqrt{2}}(|0\rangle_i \langle 0|- |1\rangle_i \langle 1|), \quad \hat{T}^{(i)^{\dagger}}_{1,-1}=|1\rangle_i \langle 0|\quad\\&
\hat{T}^{(i)^{\dagger}}_{1,1}=-|0\rangle_i \langle 1|,\quad |1\rangle= |\frac{1}{2}, \frac{1}{2}\rangle, \quad |0\rangle=  |\frac{1}{2}, \frac{-1}{2}\rangle.
\end{split}
\end{equation*}
After some straightforward calculation, one can obtain the Wigner distribution function of Alice-Bob subsystem as,
\begin{equation}
	\begin{split}
	\mathcal{W}_{\hat{\rho}}(\theta,\phi)&=2\pi \bigg[ |C^n_1(t)|^2 \varLambda^2_{11}+  \big(|C^n_2(t)|^2 +|C^n_3(t)|^2\big) \varLambda_{11}\varLambda_{00} + |C^n_4(t)|^2 \varLambda^2_{00}+\\&+\bigg(( C^{n^*}_2(t)+ C^{n^*}_3(t))C^{n+1}_1(t)\varLambda_{11}\varLambda_{01}+C^{n}_2(t) C^{n^*}_3(t)\varLambda_{01}\varLambda_{10}+\\&+
	(C^{n+1}_3(t) +C^{n+1}_2(t)  (t)) C^{n^*}_4 (t) \varLambda_{00}\varLambda_{01}+C^{n+2^*}_1(t) C^{n^*}_4 (t)\varLambda^2_{01}+h.c.
	\bigg) \bigg]
	\end{split}
\end{equation}
where the functions $\varLambda_{ij}$ are defined by the following spherical harmonics,
\begin{equation*}
\begin{split}
&\varLambda_{11}=\frac{1}{\sqrt{2}}(Y_{0,0}(\theta,\phi)-Y_{1,0}(\theta,\phi)),\quad
\varLambda_{01}=-Y_{1,1}(\theta,\phi),
\\&
\varLambda_{10}=Y_{1,-1}(\theta,\phi),\quad
\varLambda_{00}=\frac{1}{\sqrt{2}}(Y_{0,0}(\theta,\phi)+Y_{1,0}(\theta,\phi)).
\end{split}
\end{equation*}

\begin{figure}[!h]
	\centering
	\includegraphics[width=0.45\linewidth, height=5cm]{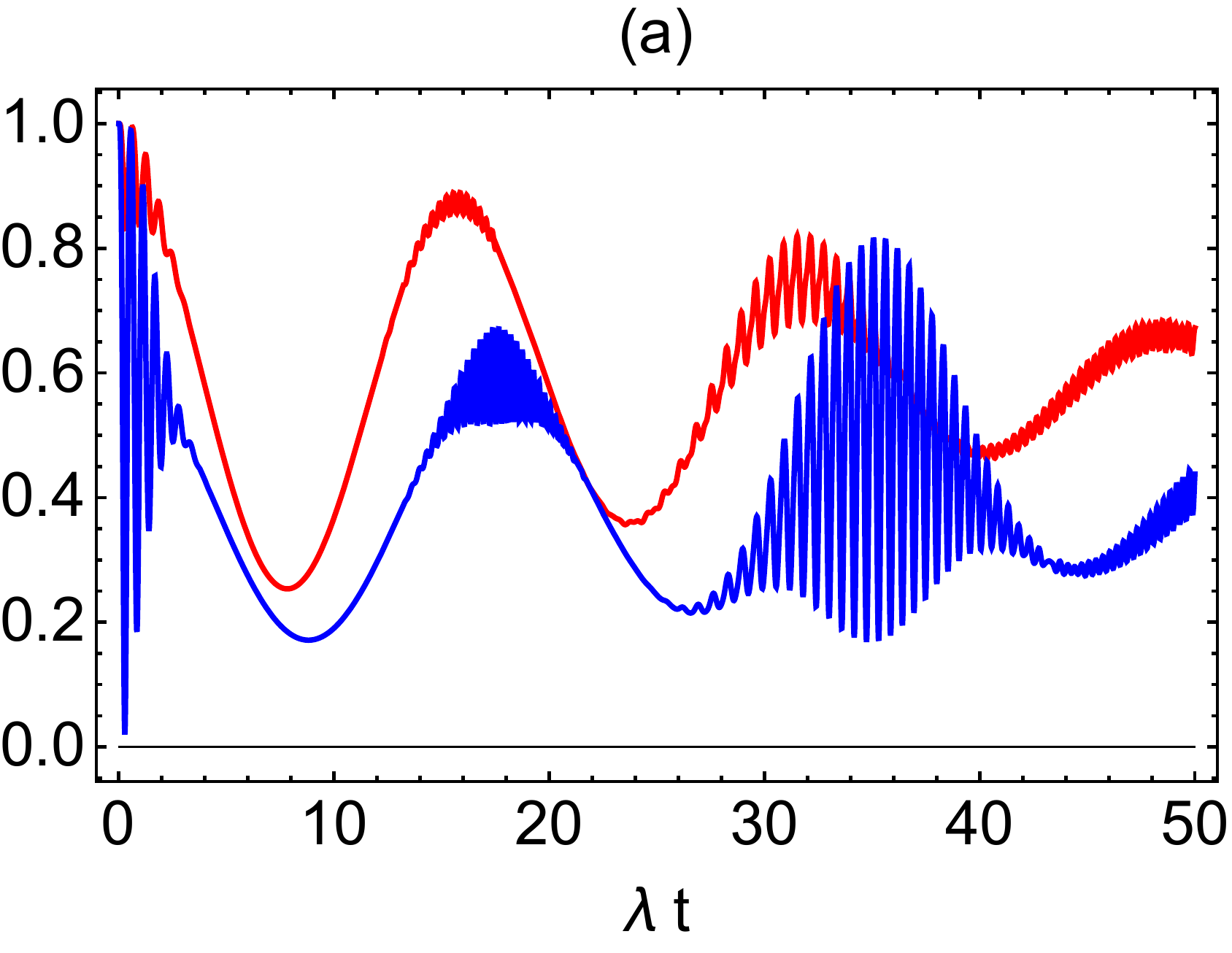}
	\includegraphics[width=0.45\linewidth, height=5cm]{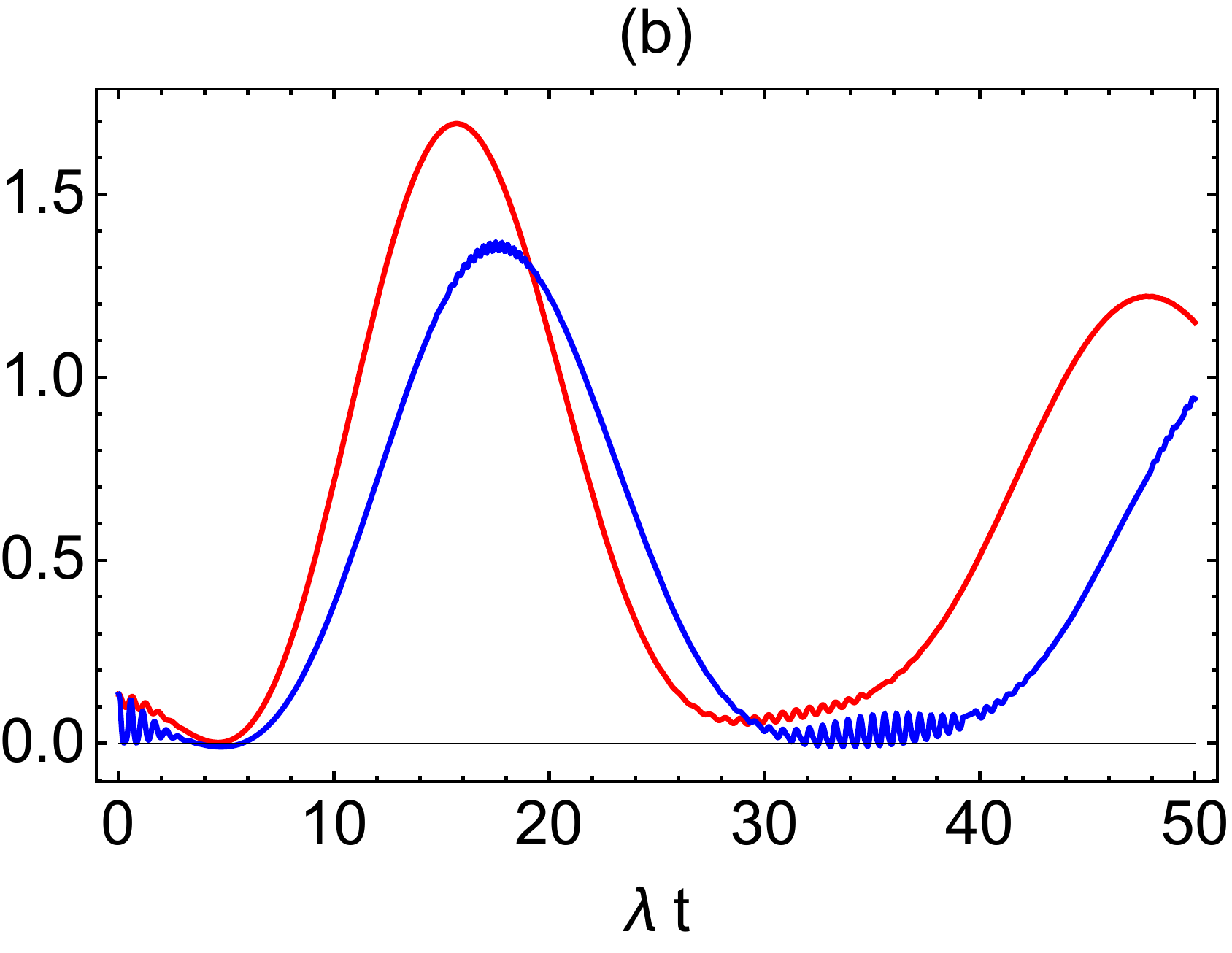}
	\caption{  Wigner distribution function against the  scaled time $\lambda t$, where there  is  no deformation, i.e., $f(n)=1$. where $\theta=\pi/2 $, $ \phi=\pi$ and $\delta =1$, $\kappa_d =1,5$ for red and blue curve, respectively.  (a) The atomic system is initially prepared in a maximum entangled state $\ket{\psi_2}=\frac{1}{\sqrt{2}}(\ket{11}+\ket{00})$. (b) The atomic system is initially prepared in a product state $\ket{\psi_1}=\frac{1}{2}(\ket{0}+\ket{1})_A(\ket{0}+\bra{1})_B$.}
	\label{fig:6.1}
\end{figure}
In Fig.(\ref{fig:6.1}), we show the behavior of the Wigner  distribution function in the presences of perfect cavity operator, namely $f(n)=1$, where we consider that the atomic system either prepared in a maximum entangled state of Bell type (Fig.\ref{fig:6.1} a) or in a product state (Fig. \ref{fig:6.1}b).  As it is displayed   from Fig.(\ref{fig:6.1}), the Wigner distribution function oscillates between its lower and upper bounds. The maximum and minimum bounds depend on the strength of the dipole-dipole interaction, $\kappa_d$. The behavior of $\mathcal{W}_p$. The phenomena of the collapse and revival behavior of the  Wigner distribution function is depicted clearly in Fig.(\ref{fig:6.1}.a), where the  atomic system is initially prepared in the maximum entangled state. On the other hand, the maximum values of the Wigner distribution function that depicted in Fig.(\ref{fig:6.1}.b), where the atomic system is initially prepared in a product state is much larger than that displayed in Fig.(\ref{fig:6.1}.a).

Our main task is investigating the behavior of $\mathcal{W}_p$, when the atomic system interacts with a deformed cavity, which means that the criterion annihilation operators are imperfect (deformed). In this context, we discuss the effect of the initial state settings of the atomic system, the initial strength of the dipole-dipole interaction, the deformation strength, and the distribution angles, on the behavior of the Wigner distribution function.
	\begin{figure}[!h]
	\centering
	\includegraphics[width=0.45\linewidth, height=5cm]{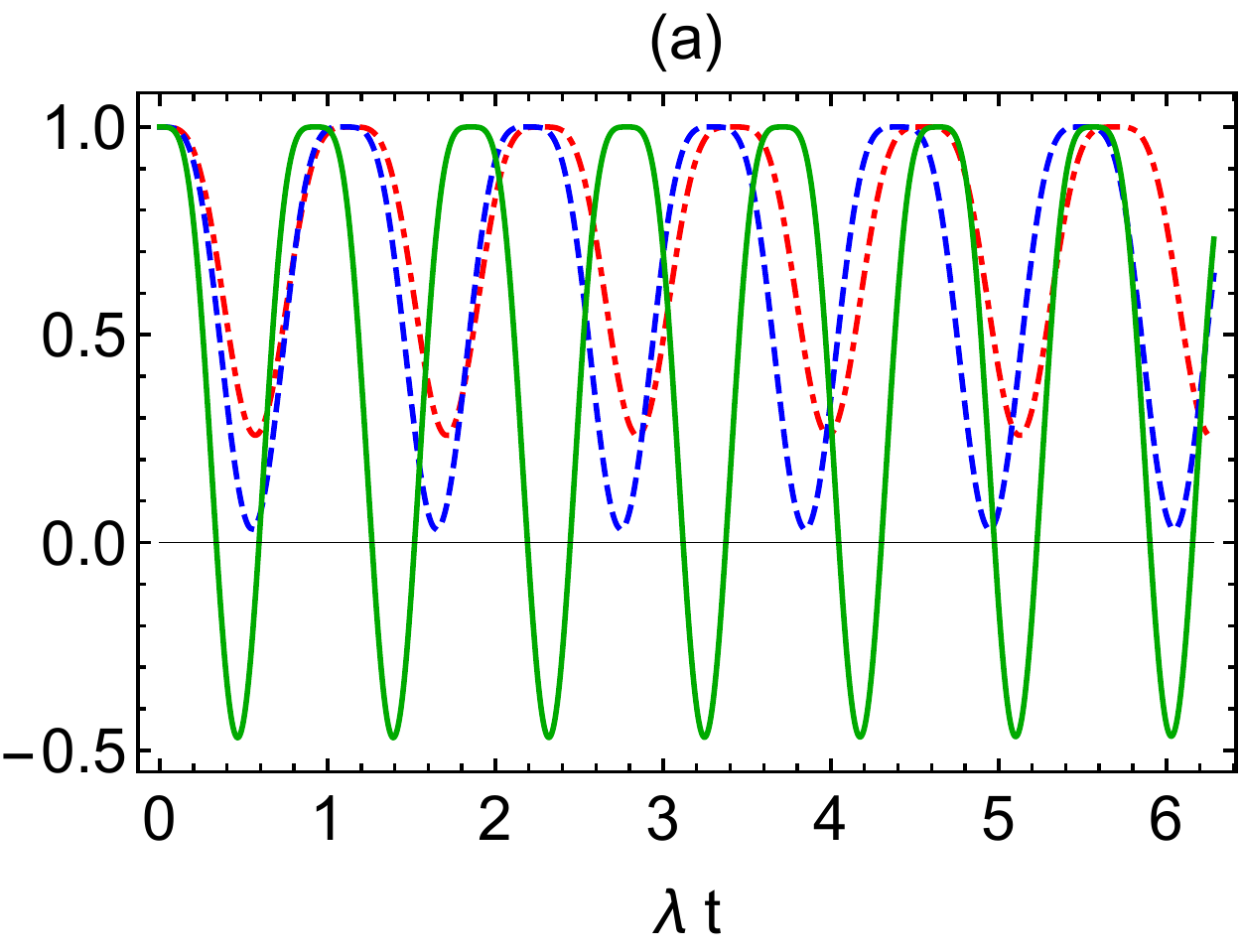}~~\quad
	\includegraphics[width=0.45\linewidth, height=5cm]{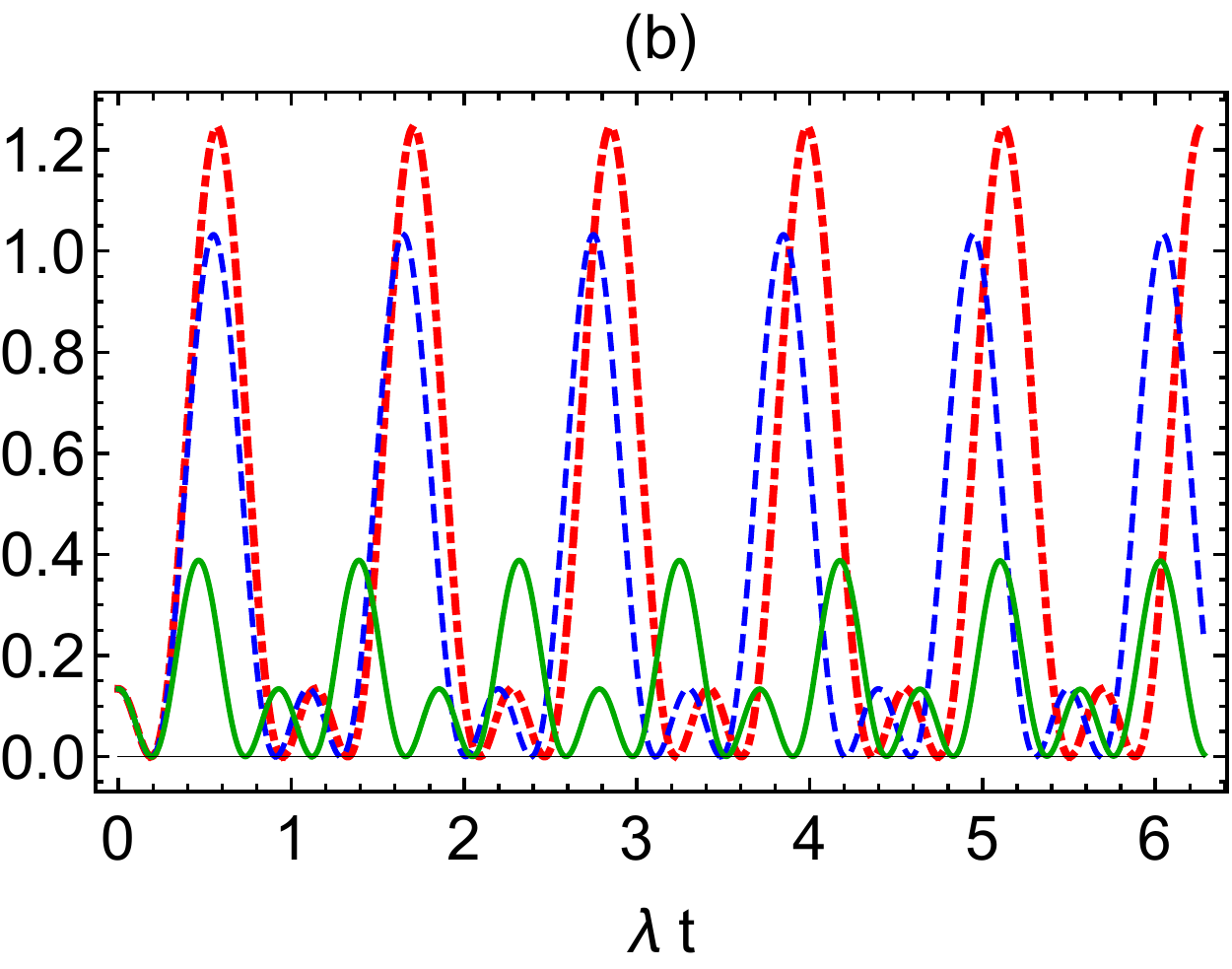}
	\caption{  Wigner distribution function against the  scaled time $\lambda t$, where there  is  no deformation, i.e.,$f(n)=\sqrt{\frac{1-q^n}{n(1-q)}}.$. $\phi=\pi , \theta=\pi/2, \delta=1, \kappa_d=5 , q= 0.1(red-dot dash),0.4(blue-dash),0.8(green-solid)$. (a) The atomic system is initially prepared in a maximum entangled state $\ket{\psi_2}=\frac{1}{\sqrt{2}}(\ket{11}+\ket{00})$. (b) The atomic system is initially prepared in a product state $\ket{\psi_1}=\frac{1}{2}(\ket{0}+\ket{1})_A(\ket{0}+\bra{1})_B$ .}
	\label{fig:6.11}
\end{figure}

In Fig.(\ref{fig:6.11}), we discuss the effect of the deformed  field operator on the behavior of the Wigner distribution function $\mathcal{W}_p$, where we assume that the atomic system is either prepared in a maximum entangled  or product state.  As it is displayed in Fig.(\ref{fig:6.11}.a), where the system is initially prepared in a maximum entangled state of Bell type, the Wigner distribution function is maximum  as $\lambda t=0$ and decreases as soon as the interaction between the atomic and the field systems is switched on. The decreasing decay depends on the strength of the deformation, where at small values of $q=0.1$ the lower bound is larger than that depicted at  large values. The behavior of the  distribution  Wigner function is repeated periodically  at further values of the interaction.  In Fig.(\ref{fig:6.11}.b), it is assumed that the atomic system is prepared in a product state. Similarly the  periodic  behavior is displayed, where $\mathcal{W}_p$ fluctuated between  its lower and maximum  bounds. However, the maximum values of the Wigner distribution is shown at small values of the deformation strength.

From Fig.(\ref{fig:6.11}), it is clear that the large deformation strength decreases the Wigner function distributing the displayed for atomic  system is initially prepared in a maximum entangled state. On the other hand, it suppress the expected increases behavior of the Wigner function distribution, where the smallest upper values of $\mathcal{W}_p$  are displayed at large values of deformation strength.

\begin{figure}[!h]
	\centering
	\includegraphics[width=0.45\linewidth, height=5cm]{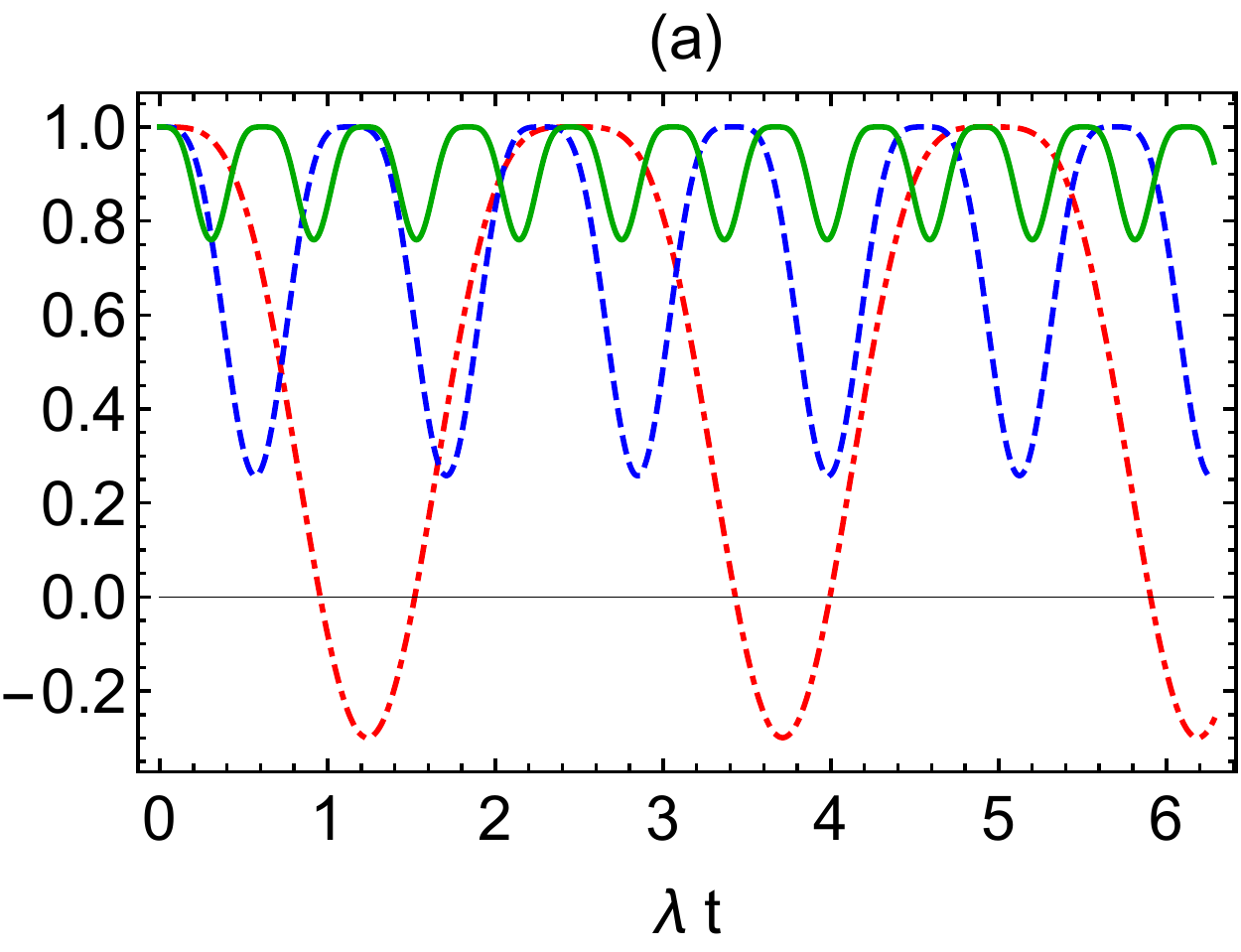}~~\quad
	\includegraphics[width=0.45\linewidth, height=5cm]{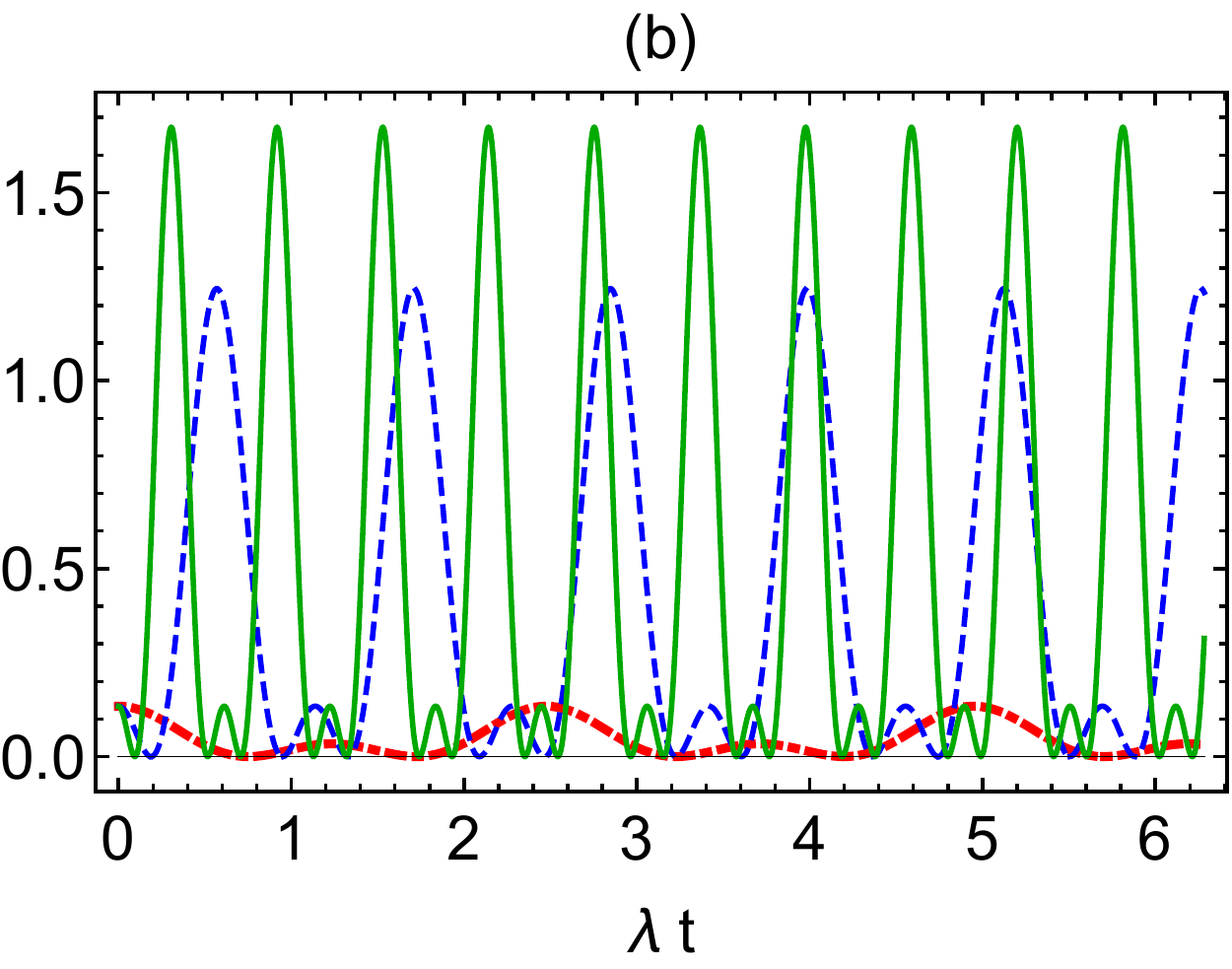}
	\caption{  The same as Fig. (\ref{fig:6.11}) with  $\phi=\pi , \theta=\pi/2, \delta=1, q=0.1 , \kappa_d$= 1(red-dot dash),5(blue-dash),10(green-solid).}
	\label{fig:6.12}
\end{figure}

In Fig.(\ref{fig:6.12}), we investigate the effect of the dipole-dipole interaction strength on the behavior of the Wigner distribution function in the presences of deformed field cavity. The behavior is similar to that displayed in Fig.(\ref{fig:6.1}), namely $\mathcal{W}_p$ oscillates  periodically between its lower and upper bounds. However, the number of oscillations and their amplitudes depends on the dipole-dipole  interaction strength, $\kappa_d$. It is clear that, at $\lambda t=0$  the Wigner Distribution function for a system is initially prepared in a MES is maximum. However as the interaction is switched on, $\mathcal{W}_p$ decreases gradually to reach their minimum values.  The decay on  the Wigner distribution can be countered by increasing the strength of the dipole-dipole interaction, where as it is displayed  from Fig.(\ref{fig:6.12}), as on increase $\kappa_d$, the  amplitudes of the oscillations are small, which means that the decay is small. However, as on decreases $\kappa_d$, the deformed cavity has a stronger effect and consequently the decay rate is large.

Fig.(\ref{fig:6.12}.b) describes the behavior of Wigner Distribution $\mathcal{W}_p$, where it is assumed that the atomic system is initially prepared in a product state. It is clear that $\mathcal{W}_p$ oscillates periodically, where their amplitudes increase as one increases the dipole strength, $\kappa_d$.  This means that the destructive effect of the deformed cavity may be Withstand   by increasing the strength of the dipole strength. On the other hand, at small values of the interaction strength $\kappa_d$, the Wigner distribution is very weak  against the deformed cavity, where their amplitudes are very small and  vanish periodically.

\begin{figure}[!h]
	\centering
	\includegraphics[width=0.45\linewidth, height=5cm]{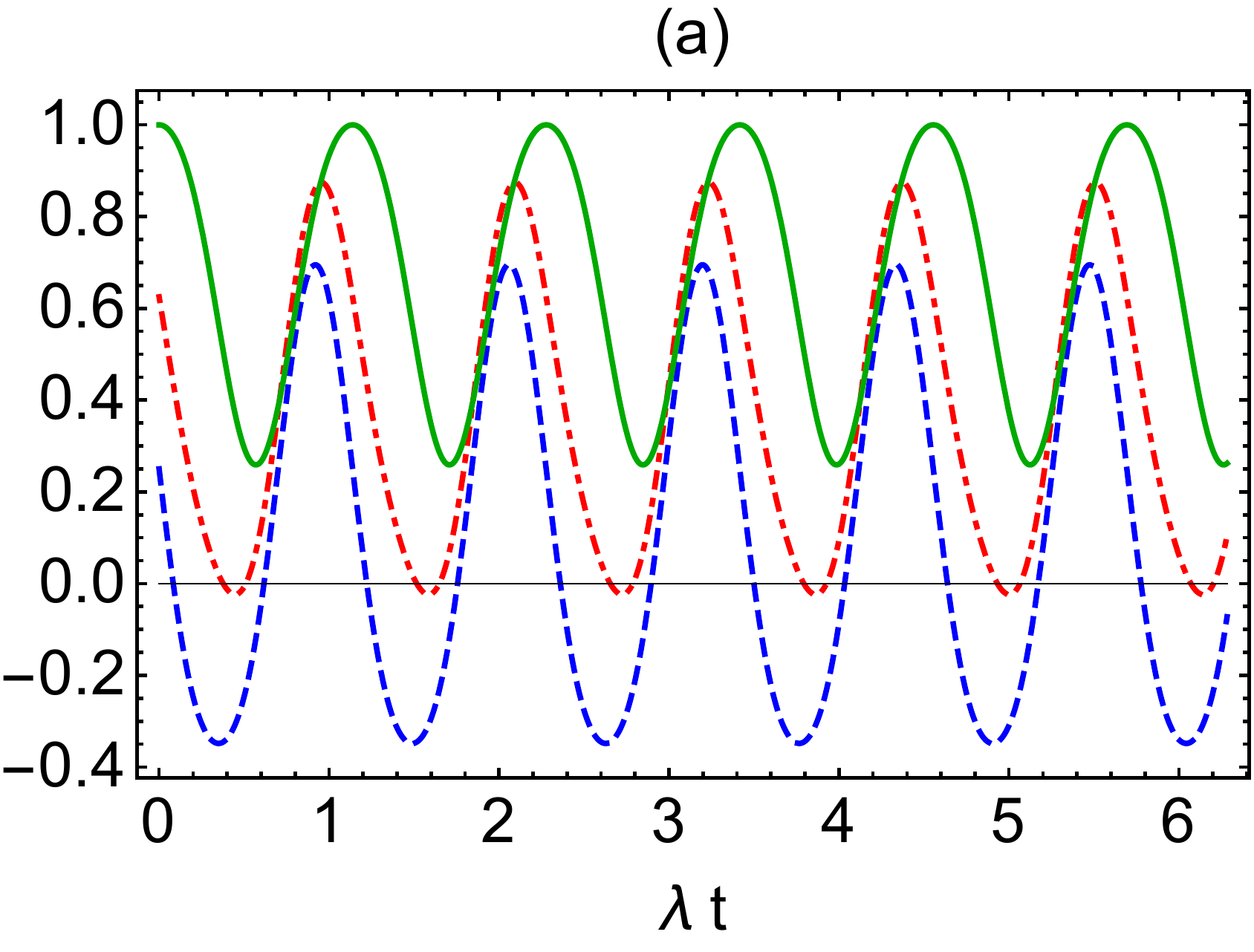}~~\quad
	\includegraphics[width=0.45\linewidth, height=5cm]{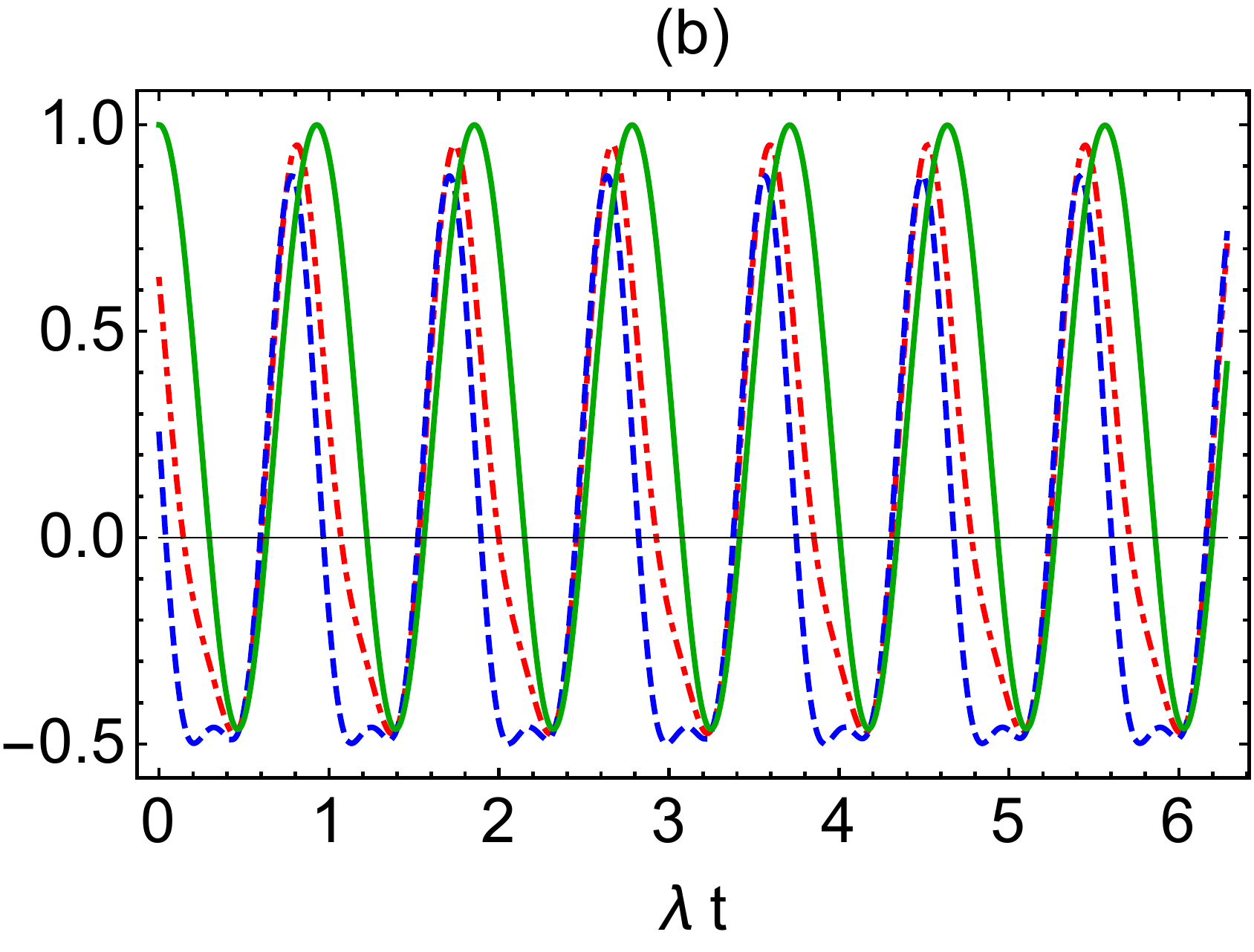}
	\caption{  Wigner distribution function against the  scaled time $\lambda t$, for atomic system is initially prepared in a maximum entangled state  $\ket{\psi_2}=\frac{1}{\sqrt{2}}(\ket{11}+\ket{00})$ where we set  $ \theta=\pi/4, \delta=1, g=5 , \phi= \pi/4 (red-dot dash),\pi/2(blue-dash), \pi(green-solid)$ (a)$q=0.1  (b)  q=0.8 $}
	\label{fig:6.p1}
\end{figure}

In Fig.(\ref{fig:6.p1}), we discuss the effect of the distribution angles $\theta$ and $\phi$ on the behavior of the deformed Wigner Function $\mathcal{W}_p$, where we set the weight angle $\theta=\pi/4$ and different values of the phase angle $\phi$. The general behavior shows that $\mathcal{W}_p$ oscillates periodically between its maximum and minimum values. It is clear that, these parameters play  as control parameter  on maximize/ minimize the Wigner distribution function.
As it is clear  from Fig.(\ref{fig:6.p1}) the amplitudes of  $\mathcal{W}_p$ increases as the deformation strength increases, and consequently the upper bounds are improved. However, different choices of the phase angle maximize the amplitudes of the Wigner distribution. On the other hand, as it is depicted from Fig.(\ref{fig:6.12}a) and (\ref{fig:6.p1}), the weight angle $\theta$ has a clear effect on  increasing the amplitudes' oscillations, hence the minimum values decreases.
 This means that by controlling on the  distribution angles, one can increase the possibility of suppressing the decoherence induced by  the deformed cavity.

	\begin{figure}[!h]
		\centering
		\includegraphics[width=0.45\linewidth, height=5cm]{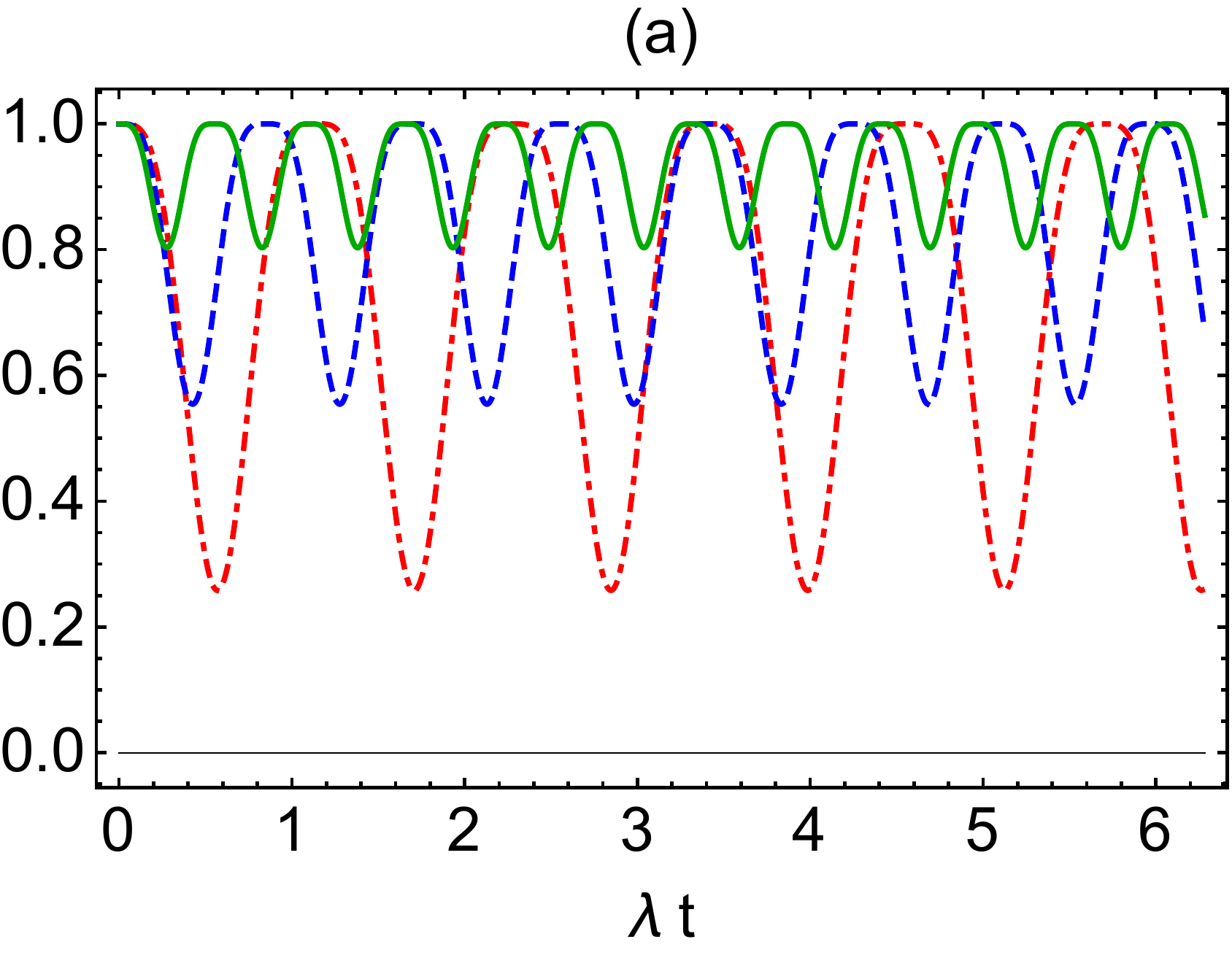}~~\quad
		\includegraphics[width=0.45\linewidth, height=5cm]{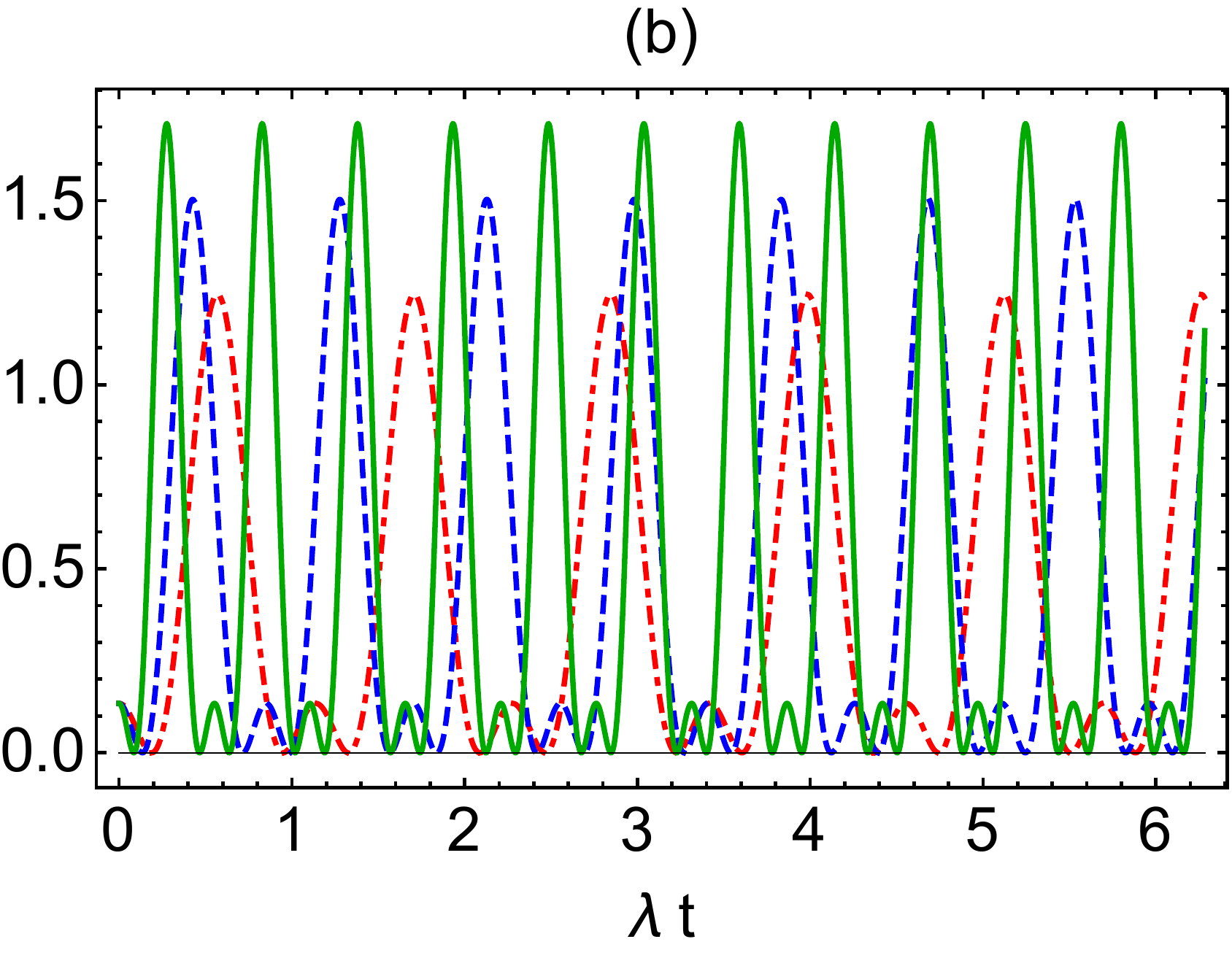}
		\caption{  The same as Fig. (\ref{fig:6.11}) with  $\phi=\pi, \theta=\pi/2, q=0.1, g=5 , \delta= $1(red-dot dash),5(blue-dash), 10(green-solid).}
		\label{fig:6.d1}
	\end{figure}

In Fig.(\ref{fig:6.d1}), we investigate the effect of the detuning parameter on the behavior of the deformed Wigner function for a system is initially prepared in a maximum entangled or product state. It is important to mention that the detuning $\delta=\Delta_1-\Delta_2$, where $\Delta_i,i=a,b$ is the frequency of each atom, namely the larger  $\delta$ the smaller  atomic frequency.   A similar behavior of,  $\mathcal{W}_p$  is displayed, where it oscillates between its maximum and lower bounds. As it is clear from Fig.(\ref{fig:6.d1}.a),  at small values of $\delta$, the decreasing rate of the Wigner function is large. However, the oscillations's amplitudes decreases as $\delta$ increases and consequently the lower bounds of the Wigner function are improved. The similar effect is predicted in Fig.(\ref{fig:6.d1}.b), where the  atomic system is initially prepared in a product state.  The maximum bounds of $\mathcal{W}_p$ are displayed at larger detuning parameter. Therefore, by controlling  the atomic  system frequency  one can  suppress the decay induced from the deformed cavity

\section{Conclusion}\label{s6.4}

A system consists of two atoms interacts locally with a deformed cavity. It is assumed that the atomic system is either prepared in a maximum entangled state of Bell state type, or in a product states. We investigate the effect of deformed cavity, which is represented by imperfect field operators, on the behavior of the Wigner distribution function .
our results, show that in the presences of perfect cavity the Wigner distribution function, the maximum and minimum bounds depend on the strength of the dipole-dipole interaction between the two atoms.  The phenomena of the collapse and revival behavior of the  Wigner distribution function is depicted clearly when the   atomic system is initially prepared in the maximum entangled state.  However, if the system is  initially prepared  in a product state the predicted Wigner distribution function is  much larger than that displaced for maximum entangled state.

It is shown that the deformed cavity has a destructive effect on the Wigner distribution function, where  it decreases as one increases the deformation strength. The decay rate depends on the initial dipole-dipole interaction strength between the atomic subsystem. The possibility of suppressing the decay induced by the deformed cavity may be increased  by  increasing  the dipole's strength. However, if the atomic system is initially prepared in a product state the upper bounds of Wigner distribution function that predicted are much larger than that displayed  when the system is prepared in a maximum entangled state.

The effect of the distribution  weight and phase angles on the behavior of the distribution Wigner function in the presences of  deformed cavity is discussed. It is shown that, these initial parameter may be considered as a control external parameters, that  maximize/ minimize the Wigner distribution function. This means that by controlling on the distribution angles, one can increase the possibility of suppressing
the decoherence induced by the deformed cavity.

We investigate the effect of the detuning parameter on the behavior of the deformed Wigner function. It is shown that,  at small values of the detuning,  the decreasing rate of the Wigner function is large. However, the oscillations's amplitudes decreases as the detuning  increases and consequently the lower bounds of the Wigner function are improved.  The similar effect is predicted  where the  atomic system is initially prepared in a product state.  The maximum bounds of the Wigner function are displayed at larger detuning parameter. Therefore, by controlling  the atomic  system frequency  one can  suppress the decay induced from the deformed cavity

In conclusion: the possibility of suppressing the decay induced by the deformed cavity may be increased by increasing the dipole's strength or the detuning parameter. The  distribution angles may be considered as a control external parameters, that maximize/minimize the Wigner distribution function.

\end{document}